\documentstyle[12pt]{article}
\newcommand{\A}{\alpha}  \newcommand{\D}{\delta}
 \newcommand{\E}{\eta} 
 
 \newcommand{\T}{\theta}
 
\newcommand{\eps}{\epsilon} \newcommand{\del}{\partial}

\newcommand{\bea}{\begin{eqnarray}} \newcommand{\be}{\begin{equation}}
\newcommand{\ee}{\end{equation}} \newcommand{\ena}{\end{eqnarray}}
\newcommand{\beano}{\begin{eqnarray*}}
\newcommand{\enano}{\end{eqnarray*}}
\newcommand{\hf}{\frac{1}{2}} 
\newcommand{\thf}{\frac{3}{2}}
                 
\newcommand{\nn}{\nonumber \\ } \newcommand{\pa}{\partial}
\newcommand{\norm}[1]{{\protect\normalsize{#1}}} \newcommand{\LAP}
{{\small E}\norm{N}{\large S}{\Large L}{\large A}\norm{P}{\small P}}

\newcommand{\vs}[1]{\rule[- #1 mm]{0mm}{#1 mm}}

 \newcommand{\cg}{\mbox{$\cal{G}$}}
\newcommand{\ch}{\mbox{$\cal{H}$}} \newcommand{\cw}{\mbox{$\cal{W}$}}

 \newcommand{\cs}{\mbox{$\cal{S}$}}

 \newcommand{\SWZW}{ S_{{\rm WZW}}
} \newcommand{\WZW}{ {\rm WZW} } 
\newcommand{\tsl}{{sl(2)}^{\bf c}}
\newcommand{\tcg}{{\cal G}^{\bf c}}
\newcommand{\tch}{{\cal H}^{\bf c}}
\newcommand{\R}{\mbox{\rm\hspace{.0em}\rule{.042em}{.694em}\hspace{-.38em}
R$\,$}}
\newcommand{\C}{\mbox{\rm\hspace{.0em}\rule{0.042em}{.674em}\hspace{-.642em}
C$\,$}}

\def\ord#1#2{{#2\over (z-w)^{#1}}} \def\ordo#1{{#1\over z-w}}
\def\ope#1#2{#1(z)#2(w)} \def\sope#1#2{#1(Z)#2(W)}

\def\sordt#1#2{{{\T-\E}\over (Z-W)^{#1}}\left \{ #2 \right \} }
\def\sordo#1{{ \left\{#1\right\}\over Z-W}}
\def\sordot#1{{{\T-\E}\over {Z-W}}\left \{ #1 \right \} }
\newcommand{\NPB}[1]{{\it Nucl. Phys.} {\bf B#1}}
\newcommand{\PLB}[1]{{\it Phys. Lett.} {\bf B#1}}

\newcommand{\CMP}[1]{{\it Comm. Math. Phys.} {\bf #1}}

\begin{document}
\renewcommand{\thefootnote}{\fnsymbol{footnote}} \newpage
\pagestyle{empty} \setcounter{page}{0}


\hbox to \hsize{\vbox{\hsize 8 em {\bf \centerline{Groupe d'Annecy} \
\par \centerline{Laboratoire} \centerline{d'Annecy-le-Vieux de}
\centerline{Physique des Particules}}}
\hfill
\hfill \vbox{\hsize 7 em {\bf \centerline{Groupe de Lyon} \ \par
\centerline{Ecole Normale} \centerline{Sup\'erieure de Lyon}}}}
\hrule height.42mm

\vs{10}
\begin{center}

{\LARGE {\bf \protect{Real Forms of Non-abelian Toda Theories} \\[5mm]
\protect{and their \cw-algebras}}}\\[1cm]

\vs{2}

{\large \protect{J.M. Evans\footnote{e-mail:
J.M.Evans@damtp.cam.ac.uk}}}, {\em DAMTP, University of
Cambridge\footnote{Silver Street, Cambridge CB3 9EW, UK.}} \\[2mm]
{\large and} \\[2mm] {\large \protect{J.O. Madsen\footnote{e-mail:
madsen@lapphp.in2p3.fr}}}, \protect{ {\em Laboratoire de Physique
Th\'eorique }\LAP\footnote{URA 14-36 du CNRS, associ\'ee \`a l'Ecole
Normale Sup\'erieure de Lyon et \`a l'Universit\'e de Savoie}, Groupe
d'Annecy\footnote{Chemin de Bellevue BP 110, F-74941 Annecy-le-Vieux
Cedex, France.}.}

\end{center}
\vfill

\centerline{ {\bf Abstract}}

\indent We consider real forms of Lie algebras and embeddings of
$sl(2)$ which are consistent with the construction of integrable
models via Hamiltonian reduction. In other words: we examine possible
non-standard reality conditions for non-abelian Toda theories. We
point out in particular that the usual restriction to the maximally
non-compact form of the algebra is unnecessary, and we show how
relaxing this condition can lead to new real forms of the resulting
\cw-algebras.  Previous results for abelian Toda theories are
recovered as special cases. The construction can be extended
straightforwardly to deal with $osp(1 \vert 2)$ embeddings in Lie
superalgebras.  Two examples are worked out in detail, one based on a
bosonic Lie algebra, the other based on a Lie superalgebra leading to an
action which realizes the $N=4$ superconformal algebra.

\vfill \rightline{hep-th/9605126} \rightline{DAMTP/96-49}
\rightline{\LAP-A-592/96} \rightline{May 1996}

\newpage \pagestyle{plain} \setcounter{footnote}{0}
\renewcommand{\thefootnote}{\arabic{footnote}}

\section{Introduction}

Over the last twelve years there has been intensive effort devoted to the
study of two-dimensional, conformally-invariant quantum field
theories.  The tractability of these models relies on the fact that
they contain infinite-dimensional chiral symmetry algebras which
organize the states or operators into manageable numbers of
representations. Such chiral algebras always contain the Virasoro
algebra, but they may also incorporate supersymmetries, Kac-Moody
currents, or other, higher-spin, quantities. The generators of these
algebras typically do not close onto linear combinations of
themselves, but rather onto non-linear expressions, and symmetries of
this general type have come to be known as \cw-algebras \cite{BS}.

There are field theory models which realize \cw-symmetries: these are
the so-called {\it reduced WZW models\/} or {\it non-abelian Toda
theories\/}.  They are constructed using a version of Hamiltonian
reduction \cite{DrSo,BaTjDr} which can be implemented at the Lagrangian level
by gauging a WZW model \cite{ORaf} (whereas the original construction
of Drinfeld and Sokolov can be applied directly to a Kac-Moody
algebra, without ever mentioning a Lagrangian).  The construction
results in actions of the general form \be
\label{eq1}
S(g) = { 1 \over \kappa } \left [ \SWZW (g) - \int d^2z \, {\rm Tr} \{ M_+ g
M_- g^{-1} \} \right ] \, .\ee 
$G$ is some real Lie group with
corresponding real Lie algebra $\cg$; the field $g$ takes values in
some subgroup $G_0 \subset G$ with corresponding Lie algebra $\cg_0$;
$M_{\pm}$ are specially chosen elements of $\cg$; and $\kappa$ is a
coupling constant.  The first term is the usual WZW action \cite{WIT}
\be
\label{eq2}
\SWZW (g) = {1 \over 2} \int d^2 z \, {\rm Tr} ( \del g \bar \del g^{-1}) + {1
\over 3} \int_D {\rm Tr} (g^{-1} d g )^3 \ee 
in which two-dimensional space-time with (complexified) coordinates
$(z, \bar z)$ is regarded as the boundary of a three-dimensional disc
$D$. This is invariant under a pair of commuting
$\hat{\cg}_0$ Kac-Moody algebras, but the potential term involving
$M_\pm$ may break some or all of this symmetry, so that the theory as
a whole has only some smaller Kac-Moody invariance (at least as far as
the linearly realized symmetries are concerned).  In the case when the
subalgebra $\cg_0$ is abelian one recovers, for certain choices of
$M_{\pm}$, the `standard' or `abelian' Toda theories based on the
algebra $\cg$.  The ways in which $\cg_0$ and $M_\pm$ can be chosen for
a given $\cg$ will be discussed in more detail below.

The specific issues we shall consider in this paper are: the freedom that is
available in choosing different real forms for the groups $G$ and
$G_0$, or algebras $\cg$ and $\cg_0$, appearing in this construction;
and the effect that these choices have on the \cw-algebra arising in
the model.  In the standard presentations of reduced WZW theories
(see \cite{ORaf} and references given there) 
attention has been confined almost exclusively to the case
in which $\cg$ is chosen to be a maximally non-compact or split Lie
algebra.  The issues we shall address arise from the
possibility of taking various other real forms for $\cg$.  It is
perhaps not widely appreciated that these additional possibilities
exist, but in fact it is not difficult to see that the standard
arguments for integrability \cite{ORaf,LS} can be extended to cover
these circumstances, and they are certainly encompassed by the most
general reduction schemes set out in \cite{ORaf}.  The new integrable
theories that emerge are interesting from a number of points of view.

First, it is natural to anticipate that choosing different real forms
for $\cg$ will lead to different real forms of the resulting
\cw-algebra, and we shall see below that this is indeed the case.  By
comparison with finite-dimensional Lie algebras, one expects these
real forms to differ in some profound ways---as regards their
representation theory, for example. Such issues have recently begun to
be investigated in the literature for the simpler case of {\it
finite\/} \cw-algebras \cite{DeB}.

Second, the construction we shall describe gives a systematic way of
finding consistent `non-standard reality conditions' for Toda
theories.  For the special case of abelian Toda theories it has
already been pointed out some time ago \cite{JE} that alternative
reality conditions are permissible consistent with integrability.  As
well as extending such results to the case of non-abelian Toda, the
work here explains the connection between these non-standard reality
conditions and Hamiltonian reduction, an issue which was never
addressed in \cite{JE}.

Third, the ideas can be immediately extended to the reduction of Lie
superalgebras, and they then play an important part in constructing
Lagrangian models with extended superconformal symmetry.  It was shown
in \cite{JE} how non-standard reality conditions are needed to obtain
Toda theories with the correct $N=2$ superconformal invariance.  We
shall see that similar considerations for non-abelian Toda models
allow the construction of a Lagrangian theory which is invariant under
the `large' $N=4$ superconformal algebra \cite{Ad}.

In the next section we discuss in more detail the class of models in
which we are interested, and the general approach to constructing new
variations of them.  The subsequent section applies these ideas to two 
examples and we conclude with some suggestions for future work.

\section{Integrable models and real forms}
\label{sec2}

There is a wide class of integrable systems which can be defined by a
Lagrangian of the form (\ref{eq1}) and the all-important property of
integrability can be established in a variety of different ways.  Thus
in \cite{ORaf} such models are obtained by gauging, or constraining
certain currents in the WZW theory associated with $G$, while the
original approach \cite{LS} makes use of Lax pairs for the
resulting partial differential equations.  The class of models we
shall consider here is by no means the most general, but it will be
sufficiently broad to explain our ideas and derive some new results.

Following closely the approach set out in \cite{ORaf} and \cite{DeRaSo,DFRS}
we can define a conformally-invariant field theory of the form
(\ref{eq1}) by specifying two things: \hfil \break
\noindent {\bf (i)} {\em An embedded} $sl(2)$ {\em subalgebra of} $\cg$.  It is
crucial to understand that we mean here an embedding of the {\it
real\/} Lie algebra $sl(2) \equiv sl(2, \R)$ into the {\it real} Lie
algebra $\cg$ (the relationship with complex embeddings will be
discussed shortly).  Let the generators of this $sl(2)$ subalgebra be
$M_0 , M_{\pm} $ obeying $[M_0 , M_\pm ] = \pm M_\pm$, $[M_+ , M_- ] =
M_0$. The eigenvalues of $M_0$ must be integers or
half-integers, in which cases we say that the embedding is integral or
half-integral respectively.  The algebra $\cg$ can be decomposed into
finite-dimensional irreducible representations of this embedded
$sl(2)$, each labeled by its spin, $j$, in the usual way.  \hfil
\break
\noindent {\bf (ii)} {\em A compatible, integral and non-degenerate grading
of} $\cg$. By a compatible grading we mean a choice of some generator
$H= M_0 + Y$ in $\cg$ such that $Y$ commutes with the entire embedded
$sl(2)$: $[ Y , M_0]=[Y , M_\pm ] = 0$.  Under the adjoint action of
$H$, the algebra $\cg$ decomposes into eigenspaces $\cg= \oplus_n
\cg_n$ labeled by their eigenvalues $n$.  The condition for an
integral grading is that these eigenvalues are integers, irrespective
of whether the $sl(2)$ subalgebra is defined via an integral or
half-integral embedding. We can now distinguish the zero-grade part
$\cg_0$ which has the special property of being a subalgebra of $\cg$
and to which we can therefore associate a corresponding subgroup $G_0$
of $G$.  When $\cg$ is decomposed into irreducible representations of
the embedded $sl(2)$, each of which has a certain spin $j$, the
generator $Y$ must be constant on each such irreducible
representation, with eigenvalue $y$, say.  The condition for a
non-degenerate grading is that this eigenvalue should not exceed the
spin: $|y| \leq j$ on each irreducible representation \cite{DeRaSo}.
\hfil \break
\noindent The two pieces of data {\bf
(i)} and {\bf (ii)} provide all the ingredients necessary to define an
integrable, conformally-invariant model of type (\ref{eq1}).  Our task
now is to characterize such embeddings and gradings in a way that is
concrete enough to allow us to calculate the resulting Lagrangians
explicitly.

We have already emphasized that the construction rests on finding
suitable embeddings of {\it real\/} Lie algebras. Nevertheless, it is
very convenient to approach this via the corresponding problem for
{\it complex\/} Lie algebras. To avoid any possible confusion, we
shall always use a superscript `{\bf c}' 
to denote a complex Lie algebra. Thus, we
shall write $sl(2)$ to mean the real Lie algebra $sl(2, \R)$ (as
introduced above) and $\tsl$ to mean the complex Lie algebra $sl(2,
\C)$. More generally, given a real Lie algebra $\cg$, its
complexification will be denoted $\tcg$.

The theory of embeddings of $\tsl$ into a complex Lie algebra $\tcg$
is very well-established and dates back to Dynkin \cite{Dy}. 
Let us introduce a Cartan-Weyl basis $t_a$ for
$\tcg$ which consists of Cartan generators $H_i$ 
together with step operators $E_{\alpha}$ for each root $\alpha$.
These obey $[ H_i, E_\alpha ] = (\alpha_i \cdot \alpha) E_\alpha$
and $[E_{\alpha_i} , E_{-\alpha_j}] = \delta_{ij} H_i$ where
$\alpha_i$ are the 
simple roots. 
There is an essentially unique {\it principal\/} embedding in any
$\tcg$, in which $M_{\pm}$ are taken to be linear combinations of step
operators for all the positive/negative simple roots:
\be
\label{prin}
M_0 = \sum_{i} \kappa_i H_i \, , \qquad M_\pm =
\sum_{i} \sqrt{\kappa_i} E_{\pm \alpha_i} \ee where 
$\kappa_i = \sum_j K^{-1}_{ij}$ and $K_{ij} = \alpha_i \cdot \alpha_j$.
The added significance of this type of embedding is that
all other possible $\tsl$
embeddings in $\tcg$ are given (up to a number of exceptions
for the D-type and E-type algebras) by principal embeddings in some
{\it regular subalgebra\/} $\tch \subset \tcg$, ie. a subalgebra
whose roots are a subset of the roots of $\tcg$.  

To pass from complex to real Lie algebras, we use the idea of an
automorphism $\tau$ of $\tcg$ which is {\it anti-linear\/}, meaning 
$\tau (a_1
\Phi_1 + a_2 \Phi_2) = a_1^* \tau (\Phi_1) + a_2^* \tau (\Phi_2)$ for
any $\Phi_1 , \Phi_2 \in \tcg$ and $a_1 , a_2 \in \C$, and {\it involutive\/},
meaning $\tau^2
= 1$ (also called an involutive {\em semimorphism}).  Such an
automorphism essentially corresponds to a notion of complex
conjugation on $\tcg$, and so it can be used to define a real Lie
algebra consisting of those elements which are fixed by the
automorphism: $\tau(\Phi) = \Phi$. Any real form $\cg$ can be obtained from
$\tcg$ in this way. (For more details, see eg.~\cite{Helg}.)

To elaborate on this, consider an 
element in $\tcg$ written in terms of the Cartan-Weyl basis 
introduced above:
\be
\label{gen}
\Phi = \sum_a \phi_a t_a \ee for arbitrary complex parameters $\phi_a$. 
We can define an automorphism $\tau$ by its action on the basis
$t_a$, checking that $\tau^2 =1$, and then extending to the whole complex
algebra using the property of anti-linearity.  
This means that we can specify $\tau$ by writing 
\be
\label{defT}
\tau(t_a) = \sum_b t_b \, \tau_{ba} \ee for some matrix $\tau_{ab}$. By
definition, the real form $\cg$ corresponding to $\tau$ 
consists precisely of the elements $\Phi$ in (\ref{gen}) for which \be
\label{rcond}
(\phi_a)^* = \sum_b \tau_{ab} \, \phi_b .  \ee 
There are two simple possibilities which illustrate this.
One obvious choice is
to take $\tau (t_a) = t_a$. The associated real Lie algebra is then,
by definition, made up of all combinations (\ref{gen}) in which the
parameters $\phi_a$ are real. This gives the {\em maximally
non-compact}, or {\em split}, real form. A second possibility is to
take \be
\label{comp}
\tau (H_i) = - H_i \, , \qquad \tau ( E_{\pm \alpha} ) = - E_{\mp
\alpha} \, .\ee
This results in the {\em compact} real form. These two real forms exist for
any complex Lie algebra $\tcg$ but in general there will be many
others. For examples, see \cite{Gil}.

Using this point of view, we have a natural way to construct $sl(2)$ 
embeddings (and
gradings) of a real Lie algebra $\cg$. We first consider an embedding
of $\tsl$ (and a grading) corresponding to a regular sub-algebra
$\tch \subset \tcg$. Then, using an automorphism $\tau$ to define the
real forms $\ch \subset \cg$, we need only check that the embedding
and grading are
consistent with the automorphism, in the sense that \be
\label{embed} 
\tau (M_\pm) = M_\pm \, , \quad \tau (M_0) = M_0 \, , 
\quad \tau (Y) = Y . \ee 
Notice that $\tau (t_a ) = t_a$,
giving the maximally non-compact form, always satisfies these
consistency conditions, and it is
this case that is the focus of most of the literature. By contrast,
the compact real form, given by (\ref{comp}), is 
never consistent with 
(\ref{embed}). Our
main interest here is to explore what happens for real forms which lie
somewhere between these two extremes.

The consistency conditions (\ref{embed}) imply that 
$\tau$ restricts to an automorphism of $\tcg_0 \subset \tcg$, and 
so the equations (\ref{gen})--(\ref{rcond}) apply in just the same
fashion to this subalgebra, defining a real form 
$\cg_0$ by specifying the allowed elements $\Phi$.
The corresponding group elements (in some neighbourhood of
the identity) can be written $ g = \exp \Phi$
and in specific cases the Lagrangian
(\ref{eq1}) can be calculated explicitly. Despite the modified reality
conditions defined by (\ref{rcond}), the entire
Lagrangian is real, by construction.  

\subsection{Abelian Toda theories}

The simplest possible examples arise when we consider a principal
embedding of $\tsl$ in $\tcg$.  Whatever real form $\cg$ we choose,
$\cg_0$ will be abelian, corresponding to a Cartan subalgebra of
$\cg$, which can be parameterized by a set of Toda fields
$\phi_i$, each associated with a simple root. 
If $\cg$ is maximally non-compact, then all the fields $\phi_i$ are
real.
For some algebras, however, there may be a non-trivial outer
automorphism $\tau$ which we can define by a permutation $\sigma$
of the simple roots:
$$ \tau(H_i) = H_{\sigma(i)} \, , \qquad \tau(E_{\pm \alpha_i}) = E_{\pm
\alpha_{\sigma(i)}} \, .
$$ The condition (\ref{rcond}) then implies $\phi^*_i =
\phi_{\sigma(i)}$, and we recover exactly the construction introduced
previously in \cite{JE}.

\subsection{\cw-algebras}

We recall (without justification) how the generators of
\cw-symmetry appear in the framework of Hamiltonian reduction.
Concentrating on the holomorphic sector of the original WZW model,
there is a conserved Kac-Moody current $J$ with values in $\cg$. In 
passing to the
reduced theory, the structure of this current can be fixed by choosing
the highest weight gauge.  In general we have $J_{hw} = M_- + \sum_{a
\in \Lambda} W^a t_a$, where the set $\Lambda$ specifies those 
generators $t_a$ which are highest-weight vectors 
of the embedded $sl(2)$, and $W^a$ are the desired conserved
quantities. 
By analogy with (\ref{rcond}), one 
can show that the induced reality
conditions for the \cw-algebra generators are \be
\label{wrc}
(W^a)^* = \sum_b \tau_{ab} \, W^b \ee 
where $\tau_{ab}$ is defined in equation (\ref{defT}).  
Notice that the energy momentum tensor always remains real and that 
the conformal weights of any \cw-generator are unaffected by the
choice of real form.

This makes explicit how different real forms of the \cw-algebra arise
but we should also point out that some care may be needed in interpreting these
conditions. If we regard the theory as defined on Minkowski
space (with the spatial direction compactified to form a cylinder)
then the interpretation is clear.
But if we pass via a conformal transformation 
to the complex plane, then the notions of reality, complex conjugation
etc.~are replaced by various hermiticity conditions on fields
(depending on their conformal weights) or on
their modes (see eg.~\cite{Gin}). It is this latter interpretation which will
be relevant when we write down operator product expansions in the
examples below.

\subsection{Supersymmetric models}

We have described how bosonic integrable models arise from 
embeddings of $sl(2)$ into Lie algebras; to construct supersymmetric 
integrable models we can consider instead embeddings 
of $osp(1\vert2)$ into real Lie superalgebras $\cg$ (see \cite{DFRS}
for a classification with comprehensive references; related work can
be found in \cite{EH,NM,DRS,IM,Ra}). 
The resulting non-abelian Toda theories are defined   
in $N=1$ superspace with coordinates $Z = (z , \theta)$
and derivatives $D=\theta \del_z + \del_\theta$ (and similarly in
the anti-holomorphic sector). 
The general superspace action is a natural extension of (1.1) in
which $\cs_{\WZW} (g)$ appears as an $N=1$ version of the WZW model 
\cite{DKPR,DRS}
for a super-group-valued superfield, and in which the 
superpotential takes the form ${\rm Str} \{  M_+ g M_- g^{-1} \}$ 
where $M_\pm$ are the {\it fermionic\/}
generators of $osp(1\vert2)$ defining the embedding. 
One simplifying feature of the supersymmetric case is that we can,
without loss of generality, set $Y = 0$ so that the embedding fixes the 
grading uniquely. 
Embeddings of $osp(1|2)$ are also 
classified (up to a few exceptions) 
as super-principal embeddings into regular
sub-superalgebras. 
The existence of different integrable models corresponding to
different real forms of a complex superalgebra $\tcg$ can therefore be
investigated just as in the bosonic case, by searching for
automorphisms which respect the $osp(1 \vert 2)$ embedding.

\section{Non-abelian examples}

Having discussed the general ideas and how these apply in the abelian
case, we now wish to consider a couple of non-abelian examples. 
For simplicity we fix $Y = 0$ in the bosonic case.
We can then define related 
non-abelian Toda theories by a pair $(\tcg ,
\tch)$ which specifies a $\tsl$ (or ${osp(1\vert2)}^{\bf c}$) embedding 
in $\tcg$ as a (super-)principal
embedding in a regular subalgebra $\tch$.
We denote the corresponding complex \cw-algebra by
$\cw(\tcg,\tch)$ and we shall see below how the real 
form of this algebra changes for suitable choices of automorphisms
$\tau$. We shall be dealing exclusively with classical \cw-algebras
but it is convenient, nevertheless, to present them using the language
of operator product expansions.

\subsection{A bosonic example: $(so(5)^{\bf c},so(3)^{\bf c})$}

This is the simplest example of a non-principal embedding into a
classical Lie algebra which meets all our requirements 
for the existence of a new real form (including the simplifying
assumption $Y=0$).
Let the long and short simple roots of $so(5)^{\bf c}$ be $\A_1$ and
$\A_2$ respectively. We take the $\tsl$ embedding in $so(5)^{\bf c}$ 
defined by the short root\footnote{This reduction has also been
considered in \cite{Bow} and, from a rather different point of view, in
\cite{Bil}.}; ie.~we 
choose $M_\pm = E_{\pm \A_2}$ and $M_0 = H_2$, which indeed defines an
integral grading. The zero-grade subalgebra is $\cg_0^{\bf c} = 
{so(3)}^{\bf c}\oplus {gl(1)}^{\bf c}$ with elements  
$\Phi + u M_0$, where
\be
\label{phi1}
\Phi = (\phi_- / \sqrt{2}) E_{-(\A_1 + \A_2)} + \phi_0 (H_1 + H_2) + 
(\phi_+ / \sqrt{2}) E_{\A_1 + \A_2}
\ee
parametrizes the non-abelian factor. The group-elements which 
appear in the action can be written $ g = \exp( \Phi ) \exp(u M_0) $. 

The maximally non-compact real form is $\cg = so(3,2)$ with 
$\cg_0 = so(2,1) \oplus gl(1)$. This is realized by taking $\phi_0$,
$\phi_\pm$ and $u$ to be real and the WZW part of the resulting action is 
$$ S_{\WZW}(g) = S_{\WZW}^{so(2,1)} + S_{\WZW}^{gl(1)}
= S_{\WZW}^{so(2,1)} - S_{\WZW}^{u(1)}.
$$
But an alternative real form can be defined by  
using the automorphism 
\bea
\label{aut1}
\tau(H_1) & = & - (H_1 + 2H_2) \, , \quad \tau (H_2) = H_2 \, , 
\nn 
\tau(E_{\pm\A_1}) & = & - E_{\mp(\A_1+2\A_2)} \, , \quad 
\tau(E_{\pm\A_2}) = E_{\pm\A_2} \, .
\ena
It is simple to check that this is involutive and compatible 
with the $\tsl$ embedding.
The real Lie algebra which it defines 
is $\cg = so(1,4)$ and the zero-grade subalgebra is $\cg_0 = so(3)\oplus
gl(1)$. This follows from 
(\ref{rcond}) applied to (\ref{phi1}) and (\ref{aut1}): 
\bea
\label{real1}
(\phi_+)^* = - \phi_- & \Rightarrow & \phi_{\pm} = \pm \phi_1 + i
\phi_2 ,
\nn 
(\phi_0)^* = - \phi_0 & \Rightarrow & \phi_0 = i \phi_3 , \ena 
with $\phi_i$ $(i = 1,2,3)$ real.
With this new choice of real form the WZW action is   
$$ S_{\WZW}(g) = S_{\WZW}^{so(3)} + S_{\WZW}^{gl(1)}
= S_{\WZW}^{so(3)} - S_{\WZW}^{u(1)} .
$$ 
The first term is the WZW action for the compact
real algebra $so(3)$ and as such it has positive-definite kinetic
part, in contrast to the previous case.
However, we must emphasize that for {\it either\/} real form the 
second term is the action for a single free field with
an overall {\em negative} sign. This relative minus 
sign between the two terms cannot be altered. 
The potential term can also be calculated, 
and we find 
\beano {\rm Tr} \{ M_+ gM_-g^{-1} \} & = & {2e^{-u} \over \rho^2}
\left[ \, \phi_0^2 + \phi_+ \phi_-
\cosh \rho \, \right] , \quad\quad \rho^2 = \phi_0^2 + \phi_+ \phi_- \, ,
\\ 
& = & {2e^{-u} \over \phi^2} 
\left[ \, 
\phi_3^2 + (\phi_1^2 + \phi_2^2)
\cos \phi \, \right] , \quad\quad 
\phi^2 = \phi_1^2 + \phi_2^2 + \phi_3^2 \, .\enano
It is clearly real, as we know it must be, for either choice of real form.

The classical algebra $\cw(so(5)^{\bf c}, so(3)^{\bf c} )$ contains 4
generators: the energy-momentum tensor $T$, a $U(1)$ generator $U$,
and two spin-two primary fields $V^\pm$, with $U(1)$-charge $\pm 1$.
They appear in the expression for the current 
in the highest weight gauge as follows: 
$$ J_{hw} = M_- + T E_{\A_2} + U (H_1 + H_2) + V^+ E_{\A_1+2\A_2} +
V^- E_{-\A_1} \, .
$$ 
In addition to the standard operator products involving the
energy-momentum tensor, we find 
\beano
\ope{U}{U} & \sim & \ord{2}{k} \, , \qquad 
\ope{U}{V^\pm} \, \sim \, \ordo{\pm V^\pm} \, , 
\\
\ope{V^+}{V^-} & \sim & \ord{4}{-6k^3} + \ord{3}{-6k^2U} +
\ord{2}{-4kU^2+2k^2T-3k^2\pa U} \\ & & + \ordo{2kUT - 2U^3 
+ k^2\pa T - k^2\pa^2U - 4k U \pa U} \, ,\\ 
\enano 
where $k$ is related to the coupling $\kappa$, and
on the right-hand side the argument $w$ is implied. 
For the maximally non-compact form, these generators are all
real. But for the new real form defined by $\tau$ in (\ref{aut1}), we 
follow the prescription (\ref{wrc}) and find that (in addition to $T^*
= T$) we have  
\beano 
(V^+)^* = -V^- & \Rightarrow & V^\pm = \pm V^1 + i V^2 \\ 
U^* = -U & \Rightarrow & U = i J 
\enano where the quantities $V^i$ $(i = 1,2)$ and $J$ are now real.
This results in an algebra 
\beano 
\ope{J}{J} & \sim & \ord{2}{-k} \, , \qquad 
\ope{J}{V^i} \, \sim \, \ordo{\epsilon^{ij} V^j} \, , \\
\ope{V^i}{V^j} & \sim & \ord{4}{3k^3 \delta^{ij}} -
\ord{3}{3k^2\epsilon^{ij}J} 
- \ord{2}{ \delta^{ij}(2 kJ^2 + k^2 T)  + 
\thf k^2 \epsilon^{ij}\pa J} \\
&& \qquad - \ordo{ \delta^{ij}(\hf k^2\pa T+ 2 k J\pa J) + 
\epsilon^{ij}(- kJT - J^3 + \hf{k^2}\pa^2J)} . \\ 
\enano

Notice that for the compact real form, 
the potential and the operator-product expansions are
manifestly invariant under an $so(2)$ symmetry (generated by $J$)
rather than under the larger $so(3)$ symmetry of the WZW action. This is
in keeping with a remark made in the introduction.
  
\subsection{A supersymmetric example: $({osp(4|2)}^{\bf
c}, {osp(2|2)}^{\bf c})$}

This is not the simplest example, but it has a number of
especially interesting features. 
We choose as simple roots of $osp(4|2)^{\bf c}$ 
bosonic roots $\A_1$ and $\A_2$ corresponding to 
an $so(4)^{\bf c}$ subalgebra, together with a single
fermionic root $\A_3$.  
The embedded ${osp(1|2)}^{\bf c}$ is defined by its fermionic 
generators $M_{\pm} = E_{\pm (\A_1+\A_2+\A_3)}+E_{\pm
\A_3}$, and the zero-grade subalgebra is then 
$\cg_0^{\bf c} = so(4)^{\bf c}\oplus {gl(1)}^{\bf c} = {so(3)}^{\bf c}
\oplus {so(3)}^{\bf c}
\oplus {gl(1)}^{\bf c}$.
We write a general element of this algebra in the form $\Phi + \Phi' +
uM_0$ where 
\be 
\label{phi2}
\Phi = \phi_- E_{-\A_1} + \phi_0 H_1 + \phi_+ E_{\A_1}
\, , \qquad 
\Phi' = \phi'_- E_{\A_2} - \phi'_0 H_2 + \phi'_+ E_{-\A_2} \, , 
\ee 
parametrizes the non-abelian factors. 
The super-group element appearing in the action is therefore
$
g = \exp(\Phi)\exp(\Phi')\exp(u M_0)\, . 
$

There is, as always, the option of choosing the split real form, but  
the main interest for us lies in the existence of another real form 
defined by the automorphism
\bea
\label{aut2}
\tau (H_1) = - H_1 \qquad  \tau(E_{\pm\A_1}) & = & -E_{\mp\A_1} \nonumber\\
\tau(H_2) = - H_2 \qquad \tau(E_{\pm\A_2}) & = & -E_{\mp\A_2} \\ 
\tau(H_3) = \phantom{-} H_3 \qquad \tau(E_{\pm\A_3}) & = &
E_{\pm(\A_1+\A_2+\A_3)} \nonumber \ena
It is easy to verify that this is
involutive, and we see immediately that it is
compatible with the embedded ${osp(1|2)}^{\bf c}$.
The corresponding real Lie superalgebra 
has a zero-grade subalgebra $so(4) \oplus gl(1)$. 
with compact non-abelian part and the 
appropriate reality conditions for the fields appearing in $\Phi$ and
$\Phi'$ above are just those written in (\ref{real1}) in the last
section. 

These conditions lead to a WZW action 
$$
\cs_{\WZW}(g) = \cs_{\WZW}^{so(4)} - \cs_{\WZW}^{gl(1)}
= \cs_{\WZW}^{so(4)} + \cs_{\WZW}^{u(1)}
$$ 
and it is very striking that this is positive-definite.
This was not the case in the previous bosonic example, despite the
emergence there of a compact non-abelian factor, because the
single free scalar field still appeared in the action with the wrong sign.
In this example, however, there is a super-trace involved in defining 
the WZW action on a Lie super-group, and this contributes 
exactly the additional minus sign required to make the scalar field
term positive. The potential term can also be computed:
\beano
{\rm Str} \{M_+ g M_- g^{-1} \} & = & 
4 e^{-u} 
 \left [ \, \cos \phi \cos \phi' + {1 \over \phi \phi'} 
( \phi^{\phantom{\prime}}_1 \phi'_1 + \phi^{\phantom{\prime}}_2 \phi'_2 + 
\phi^{\phantom{\prime}}_3 \phi'_3  ) \sin \phi \sin \phi' 
\, \right ] 
\enano
where we use the same notation as in 
(\ref{real1}) for the
parameters in $\Phi$ and $\Phi'$, with $\phi^2 = \sum_i \phi_i^2$ and 
${(\phi')}^2 = \sum_i {(\phi'_i)}^2$. 
Once again we confirm that the result is real. 

Let us now consider the implications for the $\cw$-symmetries of this model.
The complex algebra 
$\cw({osp(4|2)}^{\bf c}, \, {osp(2|2)}^{\bf c})$ 
is equivalent to the complex version of the so-called `large' 
$N=4$ algebra, which contains an $so(4)$ Kac-Moody symmetry \cite{DFRS,IM}. 
But, since we are working in an 
$N=1$ superspace formalism, this algebra will arise in an unusual
basis consisting of the super-stress tensor $T$, three spin-one
superfields $G^0 , G^\pm$ and three spin-half superfields $J^0 ,
J^\pm$.
In detail, these appear in the expression for the current in 
the highest weight gauge as follows:
\beano 
J_{hw} & = & M_- + J^- (E_{\A_1}+E_{-\A_2}) + J^+(E_{\A_2} + E_{-\A_1}) + 
J^0 \hf (H_2 -H_1 ) + \\
& & G^- E_{\A_1+\A_3} + G^+ E_{\A_2+\A_3} + G^0 (E_{\A_1+\A_2+\A_3}-E_{\A_3})
+ T E_{\A_1+\A_2+2\A_3}  \, .
\enano
To recover the usual generators of the $N=4$ superconformal algebra,
it is possible to expand these superfields in ordinary fields and to 
factorize the three spin-half fermions
\cite{Ra,GoSc} but we shall not pursue those details here. 
Omitting, for brevity, the standard operator products
involving the super-stress tensor, we find 
\beano
\sope{J^0}{J^\pm}& \sim & \sordot{\pm J^\pm} \hspace{20mm} 
\sope{J^0}{J^0} \,\,\sim \,\, \frac{2k}{Z-W} \\
\sope{J^+}{J^-} & \sim & \frac{k}{Z-W} + \sordot{\hf J^0} 
\enano
\beano
\sope{J^0}{G^\pm} & \sim & \sordot{\pm G^\pm} \hspace{15mm}
\sope{J^\pm}{G^0} \,\, \sim \,\, \sordot{G^\pm} \\
\sope{J^\pm}{G^\mp} & \sim & \sordot{\hf G^0}
\enano
\beano
\sope{G^\pm}{G^\pm} & \sim & - \sordot{\frac{1}{2k} DJ^\pm J^\pm} \\ 
\sope{G^0}{G^0} & \sim & \frac{2k}{(Z-W)^2} + 
\sordot{2 T - \frac{1}{2k} J^0 DJ^0} \\
\sope{G^0}{G^\pm} & \sim & - \sordt{2}{J^\pm} - 
\sordo{DJ^\pm \mp \frac{1}{2k} J^0J^\pm} - 
\sordot{\pa J^\pm \mp \frac{1}{2k} J^\pm DJ^0} \\
\sope{G^+}{G^-} & \sim & \frac{-k}{(Z-W)^2} - \sordt{2}{\hf J^0} - 
\sordo{\hf DJ^0 + \frac{1}{2k} J^- J^+} \\
& & - \sordot{T + \hf\pa J^0 - \frac{1}{2k} J^- DJ^+} 
\enano
where $k$ is related to the coupling $\kappa$, 
the `covariant difference' in superspace is $Z-W=z-w-\T\E$, 
and the argument $W$ of the operators is implied on the right hand side. 

For the reduction of the maximally 
non-compact real superalgebra, all the generators introduced above
would be real. 
When we change the real form using $\tau$ on the other hand, we find,
following the prescription (\ref{wrc}), that the reality conditions of the 
generators become:
\beano
(J^-)^* = - J^+ & \Rightarrow & J^\pm = \hf( \pm J^1 + i J^2) \\
(J^0)^* = - J^0 & \Rightarrow & J^0 = i J^3 \\
(G^-)^* = - G^+ & \Rightarrow & G^\pm = \hf( G^1 \pm i G^2) \\
(G^0)^* = - G^0 & \Rightarrow & G^0 = - i G^3 
\enano
where $J^i$ and $G^i$ $(i = 1,2,3)$ are real. 
With these definitions the operator products simplify considerably: 
\beano
\sope{J^i}{J^j} & \sim & \frac{-2k\D^{ij}}{Z-W} + \sordot{\eps^{ijk} J^k} \\
\sope{J^i}{G^j} & \sim & \sordot{\eps^{ijk} G^k} \\
\sope{G^i}{G^j} & \sim & \frac{-2k\D^{ij}}{(Z-W)^2} + \sordt{2}{\eps^{ijk} 
J^k}
+ \sordo{\eps^{ijk}DJ^k - \frac{1}{2k} J^i J^j} \\
& & + \sordot{-2\D^{ij} T + \eps^{ijk} \pa J^k - \frac{1}{2k} DJ^iJ^j}
\ .
\enano
It is this real form which is related to the standard $N=4$
superconformal algebra. 

It is interesting that both the potential and the 
operator product expansions written above are manifestly invariant
under an $so(3)$ symmetry corresponding to the currents $J^i$, 
whereas the WZW action is based on the larger algebra $so(4)$.
In this case we know that the symmetry of the whole model is actually
$so(4)$, but in the $N=1$ superspace formulation used here only the $so(3)$ 
subalgebra is linearly realized. In passing to component fields, a
second copy of $so(3)$ emerges from the currents $G^i$. 

\section{Discussion}

As we mentioned in the introduction, the study of Hamiltonian
reduction has, up until now, been confined largely to the case of 
maximally non-compact or split real Lie algebras. 
This may be due partly to the fact that 
$sl(2)$ embeddings into maximally non-compact real Lie algebras have
a particularly direct 
relationship to embeddings into complex Lie algebras, a subject which
has been extensively investigated. 
But it seems also that it may not be 
universally appreciated that Hamiltonian reduction is possible 
at all for other real forms. 
In this paper we have introduced a method which allows us to study 
systematically the Hamiltonian reduction of arbitrary real Lie algebras
(except for the compact Lie algebras, since these have no 
$sl(2,\R)$ subalgebras) and we have made a number of points,
illustrated by two examples, to explain why the resulting theories can
be physically and mathematically interesting.

On the one hand, our method can be viewed as a way of 
finding various consistent reality conditions for the fields in a 
non-abelian Toda theory. 
On the other hand, we have explained how this corresponds to carrying
out Hamiltonian reduction of a WZW model based on
different real forms of the same complex Lie algebra,
and, as a result, we have seen that it provides a technique for
studying the different real forms of 
extended conformal algebras. 
In a following article \cite{EM}, we shall consider in more detail the 
technical and mathematical 
aspects of these ideas. 
We will show in particular that all real forms of extended conformal 
algebras can be found using this method; we will also show that we can
obtain explicitly all real forms of a complex Lie algebra 
compatible with a given Hamiltonian reduction and we shall develop 
a method to classify these. 

There are a number of other interesting questions which we hope to
pursue in the future; we mention just a few. 
First, there is the detailed study of the new forms of the
\cw-algebras we have been discussing, including their representation
theory (see \cite{DeB,Hon} for some related work).
Second, 
it would be worthwhile studying more systematically 
the reduction of real Lie superalgebras, which we have just touched on
in this paper. The example we gave showed that there is the
possibility to construct actions with positive-definite kinetic parts.
Third, we have been considering exclusively the case of conformal
reductions.
It would be very interesting to extend our methods to massive 
Toda models, and to make contact with work such as \cite{HMP}, which 
implicitly contains a number of similar ideas.
Finally, all the considerations here have been classical, and it is
clearly important to confirm that these results can be transferred to the
quantum case. \\[6pt]

\noindent
{\bf Acknowledgements:} JOM wishes to thank E. Ragoucy, 
L. F\'eher and I. Tsutsui for stimulating and helpful discussions;
JME is grateful to H. Kausch for useful remarks.
The research of JME is supported by a PPARC Advanced Fellowship.

\end{document}